\documentclass[letterpaper]{article} 
\usepackage{aaai24}  
\usepackage{times}  
\usepackage{helvet}  
\usepackage{courier}  
\usepackage[hyphens]{url}  
\usepackage{graphicx} 
\usepackage{natbib}  
\usepackage{caption} 
\usepackage{algorithm}
\usepackage{algorithmic}
\usepackage{multirow}

\usepackage{newfloat}
\usepackage{listings}
\DeclareCaptionStyle{ruled}{labelfont=normalfont,labelsep=colon,strut=off} 
\floatstyle{ruled}
\newfloat{listing}{tb}{lst}{}
\floatname{listing}{Listing}
\title{Atlas of AI Risks: Enhancing Public Understanding of AI Risks}
\author{
    Edyta Bogucka\textsuperscript{\rm 1}, Sanja \v{S}\'{c}epanovi\'{c}\textsuperscript{\rm 1},
    Daniele Quercia\textsuperscript{\rm 1,2}\\
}
\affiliations{
    \textsuperscript{\rm 1}Nokia Bell Labs, Cambridge, UK\\
    \textsuperscript{\rm 2}Kings College London, London UK\\
    edyta.bogucka@nokia-bell-labs.com.com, 
    sanja.scepanovic@nokia-bell-labs.com,
    daniele.quercia@nokia-bell-labs.com
}

\usepackage{bibentry}

\begin{document}

\maketitle

\begin{abstract}
The prevailing methodologies for visualizing AI risks have focused on technical issues such as data biases and model inaccuracies, often overlooking broader societal risks like job loss and surveillance. Moreover, these visualizations are typically designed for tech-savvy individuals, neglecting those with limited technical skills. To address these challenges, we propose the Atlas of AI Risks—a narrative-style tool designed to map the broad risks associated with various AI technologies in a way that is understandable to non-technical individuals as well. To both develop and evaluate this tool, we conducted two crowdsourcing studies. The first, involving 40 participants, identified the design requirements for visualizing AI risks for decision-making and guided the development of the Atlas. The second study, with 140 participants reflecting the US population in terms of age, sex, and ethnicity, assessed the usability and aesthetics of the Atlas to ensure it met those requirements. Using facial recognition technology as a case study, we found that the Atlas is more user-friendly than a baseline visualization, with a more classic and expressive aesthetic, and is more effective in presenting a balanced assessment of the risks and benefits of facial recognition. Finally, we discuss how our design choices make the Atlas adaptable for broader use, allowing it to generalize across the diverse range of technology applications represented in a database that reports various AI incidents.
\end{abstract}

\makeatletter
\let\@makefntext@orig\@makefntext
\renewcommand\@makefntext[1]{#1}
\makeatother

\begingroup
\renewcommand\thefootnote{}
\footnotetext{Project Website and Supplementary Materials: \\ \textbf{https://social-dynamics.net/atlas}}
\endgroup

\makeatletter
\let\@makefntext\@makefntext@orig
\makeatother

\section{Introduction}

Effectively communicating AI risks to ordinary individuals enables them to make informed decisions, advocate for their rights and interests, and push for better regulations and safer practices \cite{bao2022whose}, while also limiting unrealistic AI perceptions and expectations \cite{nourani2020role, neri2020role}. Such AI risks range from biases in decision-making to effects on jobs and collective freedoms  \cite{mcgregor2021preventing}. 

In order to communicate risks of an AI technology use, the process of its specific risk discovery needs to take place. Current approaches to risk discovery, targeting AI practitioners, include harm description templates \cite{buccinca2023aha}, impact assessment reports \cite{microsoft2022Assessment, stahl2023systematicReview}, interactive risk cards \cite{constantinides2023prompts}, and databases of technical risks \cite{IBMriskAtlas, AIRiskDatabase}. Risk discovery and AI system's impact assessment are also required to comply with regulations and standards like the European Union AI Act (EU AI Act) \cite{EUACT2024} and the NIST AI Risk Management Framework (NIST AI RMF) \cite{nist2023aiRisk}.

However, effectively communicating AI risks to ordinary individuals, even those with a strong interest in technology, remains a significant challenge. Ojewale et al. (\citeyear{ojewale2024towards}) identified only two  such communication approaches among the 390 AI auditing tools they surveyed. This challenge happens because tools like the AI Incident Database \cite{spatialDatabaseView}, which are made for experts, focus too much on technical risks, making it hard for regular people to understand how AI risks affect their lives.

Our study aims to bridge this gap by first crowdsourcing the design requirements and then developing a tool that uses information visualization techniques to fulfil these requirements and communicate AI risks to ordinary individuals interested in technology. We made four contributions:
\begin{enumerate}
    \item We conducted a crowdsourcing formative study with 40 participants to identify requirements for the new risk communication tool. Using facial recognition as a case study, participants generated only 22 unique uses, paired with 18 risks, 9 mitigations, and 8 benefits. From their feedback, we derived six design requirements: \emph{multiple uses (R1)}, \emph{balanced assessment of uses (R2)}, \emph{structured uses (R3)}, \emph{reduced complexity (R4)}, \emph{broad appeal (R5)}, and \emph{engaging exploration (R6)}.  
    \item To meet design requirements \emph{R1-R2} by increasing the number and variety of generated uses, risks, mitigations, and benefits, we used a Large Language Model (LLM) and a Generative Image Model (GIM). We started by using the LLM to devise 138 facial recognition uses across application domains. We then assessed each use for its risks and benefits. For each identified risk, where applicable, we generated mitigations that could be understood by individuals regardless of their technical knowledge. We used textual data from the LLM to generate GIM illustrations for each use and evaluated the whole generated content for correctness.
    \item To clearly communicate the evaluated content to ordinary individuals interested in technology and meet the remaining design requirements \emph{R3-6}, we employed information visualization techniques. These included visual groupings, visual metaphors, narrative patterns, interactions, and aesthetic styling, all of which contributed to the development of our tool -- the Atlas of AI Risks.
    \item We evaluated the Atlas in a crowdsourcing user study with 140 participants who reflect US population in terms of age, sex, and ethnicity, and compared it against the spatial view of the AI Incident Database. We found that the Atlas met all design requirements and participants preferred it both in terms of usability and aesthetics, finding it more helpful in shaping their decision-making process about facial recognition. 
\end{enumerate}
\section{Related Work}
\label{sec:related}

Our work draws upon literature from diverse streams, which we organized into four areas as follows.

\smallskip
\noindent\textbf{Visualizing technological risks.} AI risk assessments typically focus on specific technologies, such as facial recognition \cite{moraes2021smile}, LLMs \cite{weidinger2021ethicalsocialrisks}, and generative AI \cite{barrett2023identifying}. Visualization plays a crucial role in making these technologies' input data, components, and outputs more interpretable, aiding in identifying technological risks like data biases \cite{vis4ml_review2024,inel2023collect}. For example, visualizing dataset disparities can reveal sampling biases \cite{aif360-oct-2018}, and comparing data patterns with historical data can expose historical biases \cite{IBMriskAtlas}. Comparing inputs and outputs can show inconsistencies in decision-making \cite{whatAILearns2023}, and model vulnerabilities \cite{perturber2021}. 

\smallskip
\noindent\textbf{Overlooking human-interaction and systemic risks.} 
A systematic analysis of model cards \cite{mitchell2019model}, an essential tool for documenting AI models, reveals that AI developers often emphasize technological risks related to data and models while neglecting impacts on individuals and the environment \cite{liang2024s}. This gap arises from the difficulty in predicting a wide range of risks from diverse model uses, stakeholder types, and deployment contexts \cite{boyarskaya2020overcoming, buccinca2023aha}. To address this, Weidinger et al.\ (\citeyear{weidinger2023sociotechnical}) propose assessing risks across three sociotechnical layers: capability, human interaction, and systemic impact. The capability layer evaluates risks inherent in technical features like poor model performance, the human interaction layer addresses risks from user interactions like overreliance on AI \cite{boyarskaya2020overcoming}, and the systemic impact layer considers broader societal and environmental consequences, including unequal distribution of technology's benefits and risks \cite{aif360-oct-2018}.

\smallskip
\noindent\textbf{Visualizing risks for tech-savvy individuals.} Eppler and Aeschimann \cite{Eppler2009} noted that risk visualizations mainly target tech-savvy users, utilizing quantitative tools like matrices, bow-tie diagrams, and cognitive maps. AI risk visualizations often depict model inaccuracies with confusion matrices, accuracy curves, or activation maps, and are embedded in interactive systems handling diverse, high-dimensional data for real-time evaluation \cite{shergadwala2022human,vis4ml_review2024, kwon2022rmexplorer,johnson2023does}. Despite these tools, practitioners may misunderstand visualizations from interpretability tools \cite{interpret_ml, lundberg2017unified}, overrelying on them as proof of model readiness \cite{kaur2020interpreting}. 

\smallskip
\noindent\textbf{Overlooking visualizing risks for ordinary individuals.} Effectively communicating AI risks to ordinary individuals, even those with a strong interest in technology, is challenging and requires understanding their beliefs, experiences, and media representations of AI \cite{buccinca2024towards, AINarratives2020}. Communication techniques for this group can be categorized into three areas: \emph{explaining}, \emph{relating}, and \emph{fostering} engagement with risks \cite{visWhatWorks2021}. To \emph{explain} risks, visualizations should enhance comparisons and aid understanding, especially for those with low numeracy skills \cite{visWhatWorks2021}, using techniques like visual groupings and icon arrays. To \emph{relate} risks to existing knowledge, visual metaphors like hazard labels can be effective \cite{data_hazards}, along with personalized risk presentations using individuals' own data \cite{climateChange}. To \emph{foster} engaging exploration, visualizations should employ narrative techniques, interactive features, and aesthetic styles to enhance engagement and understanding, using methods like story structures, gamified exploration, and contrasting colors \cite{narrativeViz, Lavie2004, calculatingEmpires2023}.

\smallskip
\noindent\textbf{Research gap.} Past efforts in visualizing AI risks focused on technological aspects, neglecting human interaction and systemic risks. Additionally, these visualizations were often tailored to experts and tech-savvy users. To address this gap, we developed a new tool that shows broad AI risks and uses visualization techniques to make them accessible to ordinary individuals.
\begin{figure*}[t!]
  \centering
  \includegraphics[width=\textwidth]{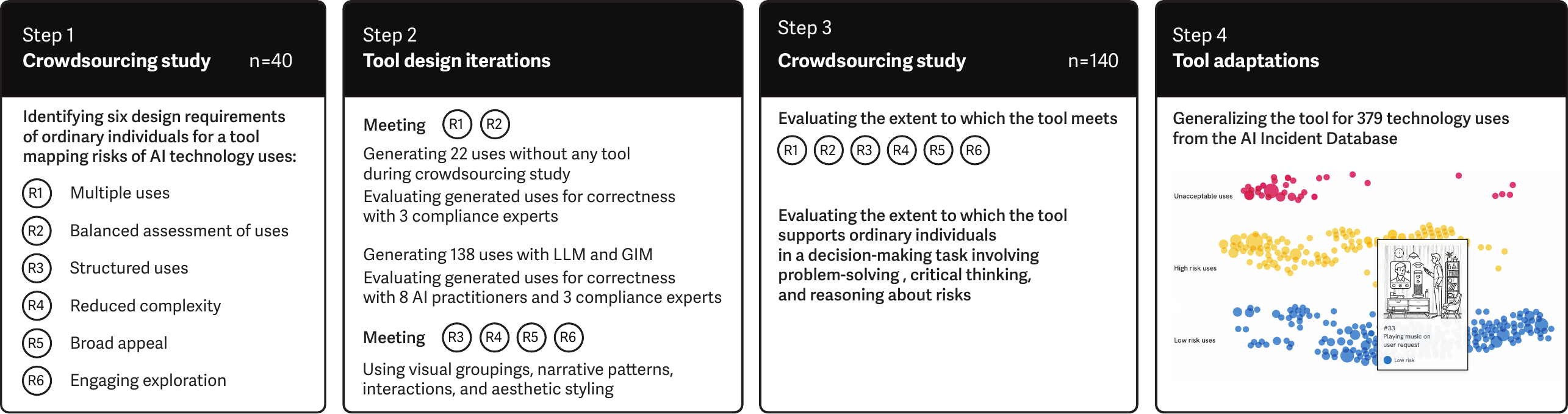}
  \caption{Proposing a tool for mapping risks of AI technology uses for ordinary individuals involved four steps. First, we conducted a crowdsourcing formative study to identify six design requirements for visualizing them. Next, using facial recognition as a case study, we asked participants of the formative study to generate its uses (including benefits, risks and mitigations) and evaluated them for correctness. Due to the low number of identified uses, we generated new uses by prompting the LLM and also evaluated them for correctness. Then, with the dataset in hand, we used information visualization techniques to build an interactive tool. We evaluated it against design requirements and its support for a decision-making task. Finally, to demonstrate how it generalizes, we visualized over 300 uses sourced from the AI Incident Database \cite{mcgregor2021preventing}.}
  \label{fig:methods}
\end{figure*}

\section{Proposing a Tool for Mapping Risks of AI Technology Uses for Ordinary Individuals}

\subsection{Identifying Design Requirements}
\label{sec:formative_study}

The formative study aimed to generate AI technology uses, identify effective design techniques for understanding their trade-offs, and establish design requirements for the tool. We selected facial recognition technology as a case study because it is well-documented \cite{facialRecognitionReview}, it is relevant for both identifying humans \cite{moraes2021smile} and animals \cite{Roberts2023_animals}, and it has sparked numerous public debates \cite{crawford2019halt}.

We recruited 40 participants interested in technology residing in the US through Prolific \cite{prolific}. These participants reflected the US population demographics \cite{census_ethnicity_2020, census_age_gender_2022} in terms of sex (21 males and 19 females) and ethnicity (24 White, 5 Black, 2 Asian, 4 Mixed or Other, and 1 Native American), with ages ranging from 19 to 63 years old. They were also digitally literate, exposed to visual media, and skilled in communication for providing feedback.

The study consisted of four parts. In the first part, we asked participants to rate on a scale from 1 to 5 (ranging from very low to very high) their knowledge in facial recognition, AI, and technology in general. In the second part, we asked them to write emails to regulators, requesting either a ban or further adoption of specific uses of facial recognition. They were also asked to explain their reasons by enumerating the risks, benefits, and steps to minimize the risks. In the third part, we presented participants with a list of 30 facial recognition uses \cite{facialRecognitionReview, Roberts2023_animals}, asking them how to best present their risks in an interactive tool. In the final part, participants were given three design techniques -- an \emph{explorative} dashboard, a \emph{narrative} infographic, and a \emph{simulative} tool with dropdowns -- and asked to choose the best one and describe its pros and cons. 

The study was approximately 30-45 minutes long and participants were paid on average about \$12 (USD) per hour. We then conducted inductive thematic analysis \cite{miles1994qualitative}, examining the uses mentioned in participant's emails and their recommendations for the tool. Participants self-reported slightly above-average knowledge in technology in general ($\mu$ = 3.8), followed by average knowledge in AI ($\mu$ = 3.2) and facial recognition ($\mu$ = 2.9). Their recommendations (quotes are marked with FP) resulted in the following six design requirements for the tool:

\vspace{1.25pt}
\noindent\textbf{(R1) Multiple uses.}
The tool should help participants to learn about a variety of uses instead of a limited number of them, as stated by FP30: \emph{``it should prompt me to consider what this technology can do''}. 

\vspace{1.25pt}
\noindent\textbf{(R2) Balanced assessment of uses.}
The tool should present each use with its risks, benefits, and mitigation strategies. This can be achieved by providing  \emph{``concrete examples''} (FP13) and \emph{`distinguishing between personal and societal risks and benefits''} (FP22).

\vspace{1.25pt}
\noindent\textbf{(R3) Structured uses.}
The tool should categorize uses for better understanding, as FP14 stated, \emph{``to help me get a clearer picture of the fields that use facial recognition''}.

\vspace{1.25pt}
\noindent\textbf{(R4) Reduced complexity.} 
The tool should present data on uses, risks, and mitigations, but its sheer volume can overwhelm users with limited technical backgrounds. To minimize this effect, the visualization should \emph{``offer different depth levels''} (FP5) and \emph{``break down the information into pieces to make it easier to come up with an opinion''} (FP21).

\vspace{1.25pt}
\noindent\textbf{(R5) Broad appeal.} 
The tool should make the uses, risks, benefits, and mitigation strategies accessible and relevant to individuals interested in technology, regardless of their technical background. As stated by FP20, \emph{``examples should relate to issues and concerns that people commonly have about AI''}. The uses should be visualized in \emph{``a less complicated way, not like designs for a technical audience''} (FP40).

\vspace{1.25pt}
\noindent\textbf{(R6) Engaging exploration.} 
The tool should engage users with \emph{``many interactive elements to allow for deeper exploration''} (FP5) or a \emph{``guided tour''} (FP28).

Over half of participants ($n$ = 22) preferred the narrative technique for the tool because it closely matched R4 by simplifying complex data into understandable snippets, and R6 by engaging users with a coherent flow of information, which helps maintain the viewer’s interest and attention. However, participants raised potential concerns that it could limit user flexibility to explore the data independently and introduce bias if it overfocuses on risks or benefits.

\subsection{Meeting the First and Second Design Requirement}
We outline the design decisions we made, informed by previous research and expert feedback, to ensure the tool effectively presents \textbf{(R1) Multiple uses} and \textbf{(R2) Balanced assessments of uses}.

\subsubsection{Generating uses without any tool.}
To generate many uses of facial recognition and identify their associated risks, mitigations, and benefits, we adopted the EU AI Act's five-component definition of use \cite{Golpayegani2023Risk} and thematically analysed the participants' emails \cite{miles1994qualitative}.

First, we identified phrases in the emails related to one of eight categories: \emph{purpose} (the AI's end goal, e.g., verifying traveler identity at border controls), \emph{capability} (technological solution behind the AI, e.g., matching faces to criminal databases), \emph{AI subject} (those impacted by the AI, e.g., travelers),  \emph{AI user} (the entity managing the AI, e.g., border control agency), \emph{domain} (the specific sector where the AI is applied, e.g., border control management), \emph{risk} (e.g., infringing on the right to privacy), \emph{mitigation} (e.g., implementing an opt-out option), and \emph{benefit} (e.g., improving security measures).
Second, we organized the phrases by category, grouping similar ones together. We summarize all the generated uses, risks, mitigations, and benefits in Supplementary Materials, Appendix A.

\subsubsection{Evaluating uses generated without any tool.}
To evaluate the correctness and variety of generated uses, we introduced four quantitative metrics. The first metric assessed the number of correct uses by implementation potential and risk level per the EU AI Act. We defined correct use as technically feasible, considering its applicability and usability, and categorized its implementation potential into existing (currently in use), upcoming (in development or early prototype stage), and unlikely (lacking applicability and usability). The second, third, and fourth metrics assessend the number of correct risks, mitigations, and benefits, defined as those that are realistic and likely to occur or succeed when implemented. We categorized these into three types—technical capability, human interaction, and systemic impact—based on the existing taxonomy of socio-technical evaluations \cite{weidinger2023sociotechnical}.

We independently assessed each generated use to first determine its correctness and implementation potential and then agreed on the final assessment. To assess the risk level of use, we recruited three AI compliance experts from our company who classified uses as unacceptable risk, high-risk, or low-risk, with justifications for each label.

Participants jointly identified 22 correct uses: 21 existing and 1 upcoming (Supplementary Materials, Appendix A). The top three uses mentioned were unlocking devices ($n$ = 8), identifying suspects for crime prevention ($n$ = 5), and apprehending individuals on the run ($n$ = 4). Four uses were considered unacceptable (e.g., tracking citizens in public spaces for law enforcement), 15 high-risk, 3 low-risk, and 1 varied between high-risk and unacceptable depending on context. Participants identified 18 correct risks, 8 correct benefits, and 9 correct mitigations, mostly focusing on systemic impacts ($n$ = 11, $n$ = 6, and $n$ = 4, respectively).

\subsubsection{Generating uses with the LLM.}
Formative study participants primarily generated high-risk uses, overemphasized risks compared to benefits and mitigations, and barely addressed risks related to technical capability and human interaction, likely due to limited knowledge of AI development. These are, however, the most common risks resulting in incidents \cite{mcgregor2021preventing}. To ensure the tool meets the two design requirements, we needed to increase the variety of generated uses, risks, mitigations, and benefits. We achieved this by using four prompts for LLM-assisted AI impact assessment — \emph{ExploreGen} \cite{herdel2024exploregen}, \emph{RiskGen}, \emph{BenefitGen}, and \emph{MitigationGen} \cite{constantinides2024_risks_benefits, AIDesign2024} — and engineering one prompt for a generative image model — \emph{IllustrationGen}. We report the prompts' content in Supplementary Materials, Appendix B.

\subsubsection{Evaluating uses generated with the LLM.}
Similarly to the previous evaluation of uses generated without any tool, we introduced five quantitative metrics to evaluate the correctness and variety of generated uses. In addition to the already used four metrics -- the number of correct risks, mitigations, and benefits by type -- we evaluated each illustration for correctness of use depiction, defined as recognizable, relatable, and free of visual stereotypes.

Each metric was evaluated using a two-step process involving two authors and external experts. For assessing correctness, implementation potential, and depiction, we first independently evaluated each generated use, then discussed and agreed on the final assessment with the research team. For risk level, we familiarized ourselves with the EU AI Act \cite{EUACT2024} and randomly sampled 18 uses. We then recruited three AI compliance experts from our company to annotate these uses with risk labels and provide justifications. Afterwards, we independently assessed the remaining 120 uses for risk level, reaching a final consensus through discussions with the research team. For the number of correct risks, mitigations, and benefits by type, we recruited 8 raters: two authors and six industry researchers and developers from our company, all experienced in responsible AI. We randomly assigned 46 uses to the raters through an interactive survey, ensuring each use was evaluated by three raters. The raters reviewed the descriptions of the uses, along with three lists of risks, benefits, and mitigations. They marked whether they agreed with each list and indicated which items should be removed if they disagreed. We then compared the correctness scores for each list.

All 138 generated uses to be correct, with 91 (66\%) already existing, 39 (28\%) being upcoming, and 8 (6\%) being unlikely. The agreement between the authors and experts and the LLM's classification was 91\%. Disagreements were about 12 uses related to environmental sustainability, agriculture, farming, and climate change mitigation, which LLM marked as existing, and the experts as unlikely. 10 uses (7\%) were identified as unacceptable, 66 (48\%) as high risk, and 62 (45\%) as limited or low risk. The agreement between the authors and experts and the LLM's classification was 90\%. Disagreements were about 14 uses, for which experts found insufficient information to derive risk label, while the LLM classified them as low risk.

By comparing the correctness scores, we found that 93\% of the risks, 95\% of the mitigations, and 82\% of the benefits were correct. The average agreements across each set of three annotators labeling the same parts of the data, as measured using the intraclass correlation coefficient, were 25\% for the risks (fair), 23\% for the mitigations (fair), and 47\% for the benefits (moderate agreement). As for the illustrations of uses, 126 (91\%) were correct, while 12 (9\%) needed re-generation due to references to national symbols ($n$ = 2), incorrect cultural depictions ($n$ = 5), and insensitive gender role representations ($n$ = 5).

\subsection{Meeting the Remaining Four Design Requirements}
\label{sec:design}

\begin{figure*}[t!]
  \centering
\includegraphics[width=\textwidth]{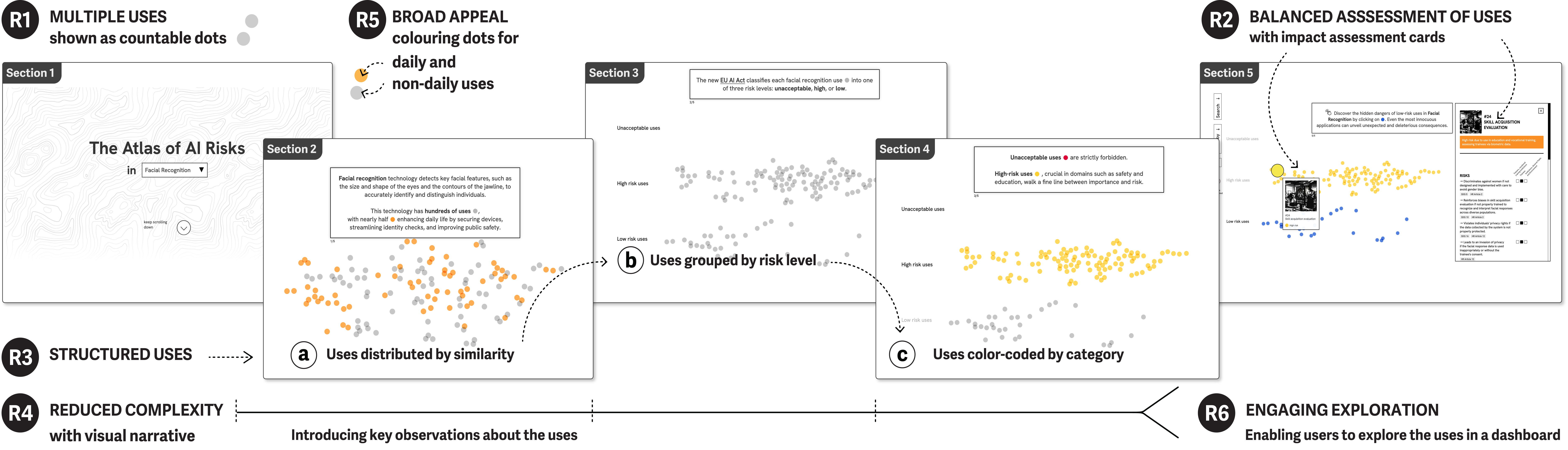}
  \caption{The interface of the Atlas of AI Risks meets six design requirements: mapping many uses of technology (R1), presenting a balanced assessment of their risks and benefits (R2) categorizing them for better understanding (R3), reducing their complexity (R4), making them relevant to ordinary individuals (R5), and making their exploration engaging (R6).}
  \label{fig:atlas}
\end{figure*}

\begin{figure}[t!]
\includegraphics[width=1\columnwidth]{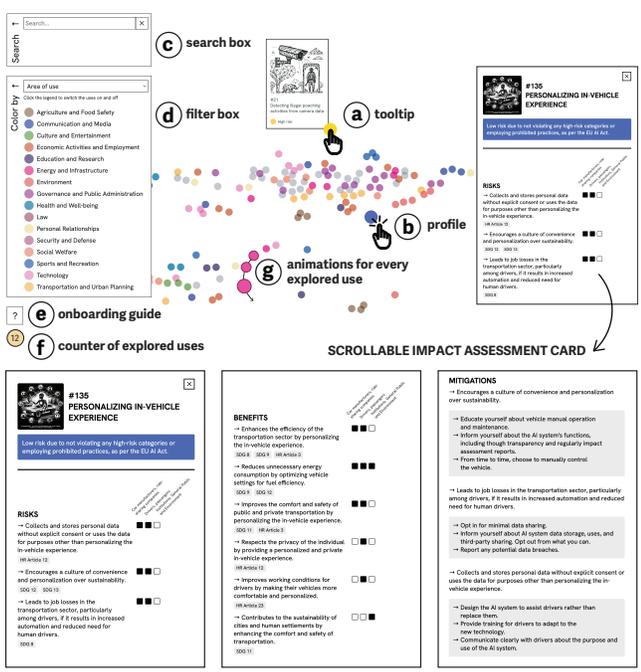}
  \caption{The final dashboard for use exploration. It includes an impact assessment card available in two versions—a brief tooltip (a) and a detailed profile (b) listing risks, benefits, and mitigations—as well as interactions for use browsing, onboarding, and exploration tracking (c-g).}
  \label{fig:dashboard}
\end{figure}

We describe the design decisions, informed by previous research on effective communication with the public, made to ensure the tool meets the remaining design requirements.

\smallskip
\noindent\textbf{(R3) Structured uses.} 
We use a map atlas metaphor, with each use as a dot in a two-dimensional space. This visually emphasizes the need to assess risks per use and helps people understand the overall probability of risks of facial recognition \cite{visWhatWorks2021}. We then adopted three methods for varying analytical skills: spatial distribution based on semantic similarity (for a continuous view), functionality to split uses by risk (for group comparison), and new data dimensions for color-coding uses (for a discrete view). First, to spatially distribute the uses (Figure \ref{fig:atlas}a), we created sentence-level BERT (SBERT) embeddings for each use description using the \textit{paraphrase-distilroberta-base-v2} model. We used SBERT's standard settings, and showed the resulting embeddings using a JavaScript-based t-distributed stochastic neighbor embedding algorithm. Second, to facilitate comparisons between the high-risk and low-risk uses, we split their dots vertically (Figure \ref{fig:atlas}b) and added the interaction to group the uses back together \cite{visWhatWorks2021}. Third, to color-code the data, we used the existing AI Harm Taxonomy for the AI Incident Database \cite{harmTaxonomy}. We manually label each use case based on its area of application, the types of affected subjects and supervising users, as well as its impacts on critical infrastructure, children, entertainment, and the public sector (Figure \ref{fig:atlas}c).

\smallskip
\noindent\textbf{(R4) Reduced complexity.} 
We adopted a frame-based Martini Glass narrative structure \cite{narrativeViz} and a progressive disclosure design \cite{progressiveDisclosure} to reveal information gradually.

First, we unfold the complexities of assessing risks through five distinct story sections (Figure \ref{fig:atlas}, R4). The first section explains the technological features of facial recognition and introduces its 138 uses, each represented by a dot. The second section highlights the dots representing daily uses of facial recognition. The third section shows how uses can vary in risk, illustrated through an animated transition that categorizes the dots into unacceptable, high, or low risk groups. The fourth section explains the common characteristics of each risk group, emphasizing through color-coding how, despite regulations, all uses can still pose harm. The final section introduces a dashboard that encourages users to explore the uses and to find ways to mitigate risks.

Second, we paired each use with an impact assessment card that comes in two versions: a brief tooltip accessible with mouseover (Figure \ref{fig:dashboard}a), and a detailed profile accessible with a click (Figure \ref{fig:dashboard}b). The tooltip includes an illustration of the use, its short description, and a tag indicating its level of risk according to the EU AI Act. The profile includes four sections, starting with a use summary box that contains an illustration, a long description, and the overall risk level. The subsequent three sections detail the benefits, risks, and mitigations, with benefits and risks grouped by technical capability, human interactions, and social impact, and checkboxes indicating who is affected.

\vspace{12pt}
\noindent\textbf{(R5) Broad appeal.} 
We used different colors to separate the uses into two groups: daily (like unlocking smartphones) and non-daily (like tracking illegal poaching). Nearly half of the dots represented daily uses, showing how facial recognition technology is part of everyday life and making the data more relatable (Figure \ref{fig:atlas}, R5). Additionally, we carefully phrased the mitigations in the impact assessment card to be understandable regardless of (non)technical background.

\smallskip
\noindent\textbf{(R6) Engaging exploration.} 
The final dashboard includes interactions for atlas browsing, onboarding, and exploration tracking (Figure \ref{fig:dashboard}). Users can browse the atlas using the search box for keyword matching (Figure \ref{fig:dashboard}c) and the filter box for color-coding dots based on ten categories (Figure \ref{fig:dashboard}d). The onboarding guide covers the interface, tooltips, profiles, search, and filtering options (Figure \ref{fig:dashboard}e). For exploration tracking, micro-interactions include a counter that changes color as more uses are explored (Figure \ref{fig:dashboard}f) and animations for already explored uses (Figure \ref{fig:dashboard}g). 
\section{Evaluating the Tool Mapping Risks of AI Technology Uses}
The goal of the study was to evaluate if our Atlas met the design requirements and how effectively it communicated the broad risks associated with the diverse uses of facial recognition to ordinary individuals. Next, we describe our study's design (i.e., metrics), setup, execution, and results.

\subsection{Metrics}
We created a task for our study that reflects how people usually handle and question AI decisions in real life, such as writing an email to challenge or praise a system\cite{contestableCars2023}. Our participants might have done similar things before, like giving feedback to a company about its smart devices or checking their town's AI city register, where local governments list technologies used in their cities \cite{cityRegister}. Specifically, we instructed participants to ``Write a brief email to the AI policymakers and ask them to stop and approve some uses of facial recognition. For each use, argue by explaining its risks or benefits''. This task also links the risks in the Atlas to three higher-order decision-making skills: problem-solving (identifying reasonable actions), critical thinking (evaluating and synthesizing information on risks, mitigation, and benefits), and reasoning (constructing logical arguments to support actions).

We outlined six questions to evaluate whether Atlas met our design criteria (R1-6) and how effectively it supported individuals in completing the task, asking, ``How successful was the tool in...    
\begin{enumerate}
    \addtolength{\leftskip}{5pt}
    \setlength{\itemsep}{0pt}
    \setlength{\parskip}{0pt}
    \setlength{\parsep}{0pt}
    \item [Q1] ...communicating multiple uses (R1)?
    \item [Q2] ...providing a balanced assessment of uses (R2)?
    \item [Q3] ...structuring uses (R3)?
    \item [Q4] ...reducing complexity (R4)?
    \item [Q5] ...achieving a broad appeal (R5)?
    \item [Q6] ...engaging individuals in exploration? (R6)?
\end{enumerate}

We then defined four quantitative and four qualitative metrics to answer these questions. 

\smallskip
\noindent\textbf{Quantitative metrics.}
The first quantitative metric assessed the extent to which the tool provided a balanced assessment of uses. It was measured as the percentage of participants who agreed with the statement: \emph{``The tool helped me to understand both the risks and benefits of facial recognition''}. Three metrics measured how successful the tool was in engaging users in exploration. 

The first metric was the about the \emph{usability} of the tool and it was measured through the System Usability Scale~\cite{brooke1996sus}. The second metric was about the \emph{visual aesthetics}, understood as classic aesthetics (for clarity and order), expressive  aesthetics (for originality), and pleasurable interaction (for user enjoyment), three factors crucial for making technology use exploration intuitive and engaging for ordinary individuals. We measured them through the Perceived Visual Aesthetics scale \cite{Lavie2004}. The third metric was about the \emph{exploration time} required for a participant to complete the task.

\smallskip
\noindent \textbf{Qualitative metrics.}
Qualitative metrics were captured through four open-ended questions. The first assessed the effectiveness in communicating multiple uses, by asking: \emph{``How successful was the tool in helping you learn about a variety of uses?''} The second measured the usefulness of structuring uses, by asking \emph{``How useful were for you the categories of uses shown in the tool?''}. The third measured effectiveness in reducing complexity by asking: \emph{``How successful was the tool in simplifying the information?''} The fourth assessed the \emph{relevance} of presented information to ordinary individuals: \emph{``How successful was the tool in identifying uses, risks, benefits, and mitigations relevant to you?''}. 

To assess the impact of knowledge on responses, participants self-evaluated whether they were more skilled or knowledgeable than the average person in the task, technology, facial recognition, and AI.
    
\subsection{Setup}
The study consisted of seven steps (Supplementary Materials, Appendix C). In the first step, participants were introduced to the first reflective judgment task and were asked to write the first email to the regulators without using any tools. In the second step, participants completed the self-assessment control questions, alongside the first randomly assigned attention-check. Moving to the third step, participants interacted with either our Atlas (treatment) or the baseline (control), analyzed the technology's uses in the visualization and completed the second reflective judgment task. After interaction with the treatment or control, participants evaluated its usability and completed the second attention-check sentence in the fourth step. In the fifth step, participants assessed the visualization's visual aesthetics. \hspace{1cm} This step also included the third attention check. In the sixth step, participants explained if the tool helped them to learn about multiple uses of facial recognition, if the proposed categories of uses were useful, and if the content presented in the visualization was relevant to them. Finally, in the last, seventh step participants provided recommendations for future development of the visualization.
\smallskip

\begin{figure}[h!]
  \centering
\includegraphics[width=\textwidth]{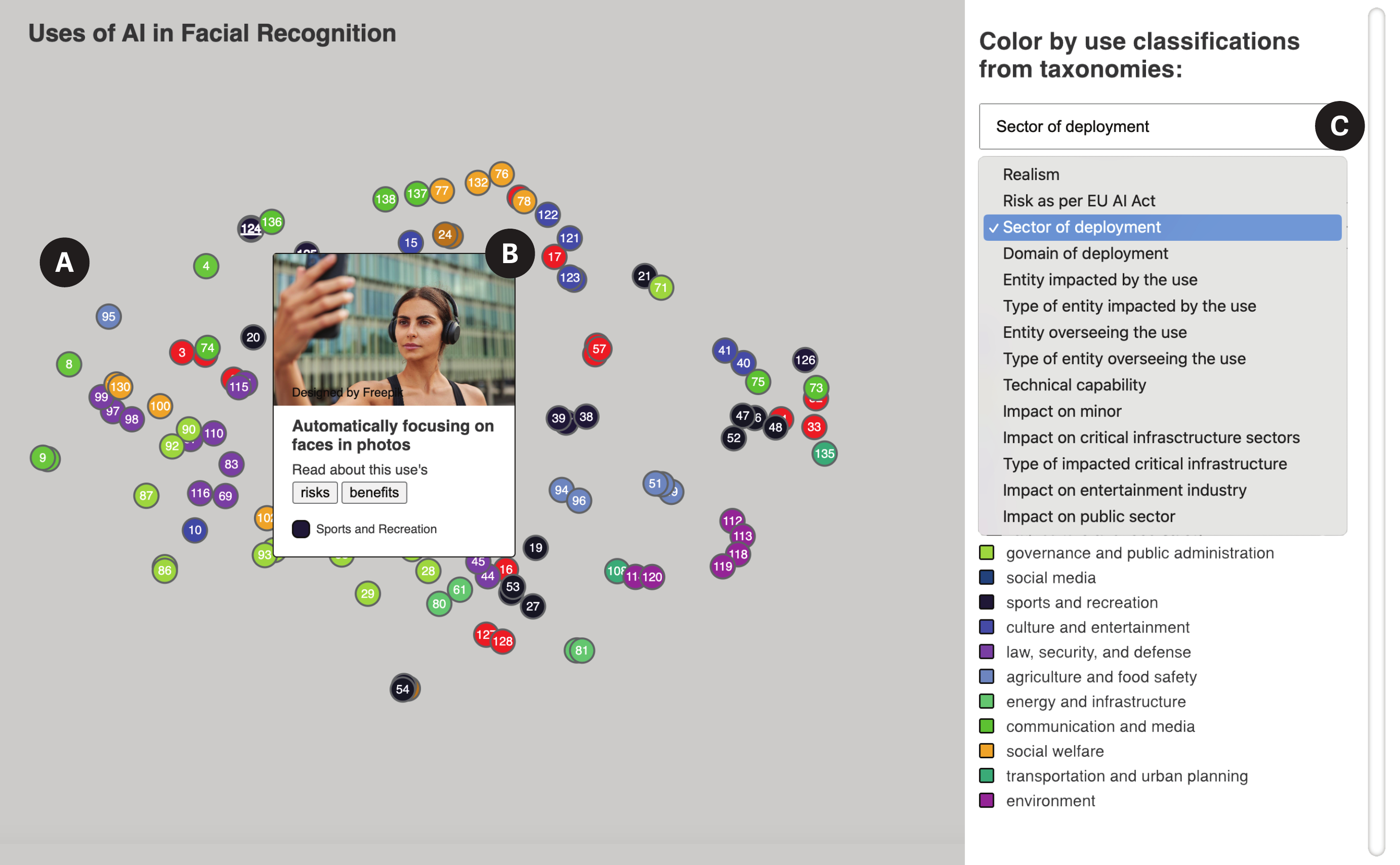}
  \caption{\textbf{The baseline visualization mirrored the state-of-the-art AI risk visualization} \cite{spatialDatabaseView}. A one-page dashboard approach groups technology uses based on their similarity scores (A). Viewers can access their descriptions and news articles about their associated risks and benefits (B), and filter them through a dropdown menu (C).}
  \label{fig:baseline}
\end{figure}

\noindent\textbf{Baseline.} 
The control group's visualization mimicked state-of-the-art AI risk visualization -- a dashboard (Supplementary Materials, Appendix D, Figure \ref{fig:baseline}). It displayed technology uses grouped by similarity on the left, pop-ups for the uses in the center, and a dropdown menu with a legend on the right. Like the Atlas, it displayed various uses (R1), categorized them (R2), and allowed exploration (R6). However, it differed in balancing risks and benefits (R2), reducing complexity (R4), and appealing to audience (R5). 

\subsection{Execution}
\label{subsec:execution}

\noindent\textbf{Participants.}
We recruited participants from Prolific~\cite{prolific}, aiming for a sufficiently large number in both the treatment and baseline groups to achieve statistical significance. The study focused on individuals interested in technology, who are representative of the general U.S. population, for two key reasons. First, recruiting English native speakers ensured comprehension of the study materials, enhancing results reliability. Second, exposing US participants to a risk-based approach to AI regulation is relevant, as both the US and EU adopt similar frameworks, with crosswalk documents showing their compatibility. Recruited study participants were compensated approx. \$12 (USD) per hour. 

\smallskip
\noindent\textbf{Procedure.}
We developed a web-based survey that included a reflective judgment task and administered it on Prolific. The survey comprised seven pages, each corresponding to a setup step plus a final confirmation screen. In step three, participants were randomly presented with a link to access either the treatment or the control in a new browser window.

To ensure response quality, we implemented four survey design features. First, we disabled pasting text from external sources and prohibited editing previous responses to encourage original answers and maintain survey flow. Second, we set word ranges for open-ended questions (50-250 words~\cite{liang2024s}) and validated them in real-time to prevent survey fatigue. Third, we used a click tracker to verify participants accessed the treatment or baseline link. Fourth, we randomly administered one of three attention checks. To be included, participants had to correctly respond to at least two attention checks and click the link.

\smallskip
\noindent\textbf{Analysis.} We performed both quantitative and qualitative analyses. 
For the quantitative analysis, we measured four metrics: the share of participants who found the visualization helpful for understanding both the risks and benefits; the average SUS usability score; the average value of perceived aesthetics across classic aesthetics, expressive aesthetics, and pleasurable interaction; and the average time to complete the task. For the qualitative analysis, we thematically analyzed the open-ended questions \cite{miles1994qualitative} to identify repeating themes within design requirements and key factors affecting decision-making.

\begin{figure}[t!]
  \centering
  \includegraphics[width=1\columnwidth]{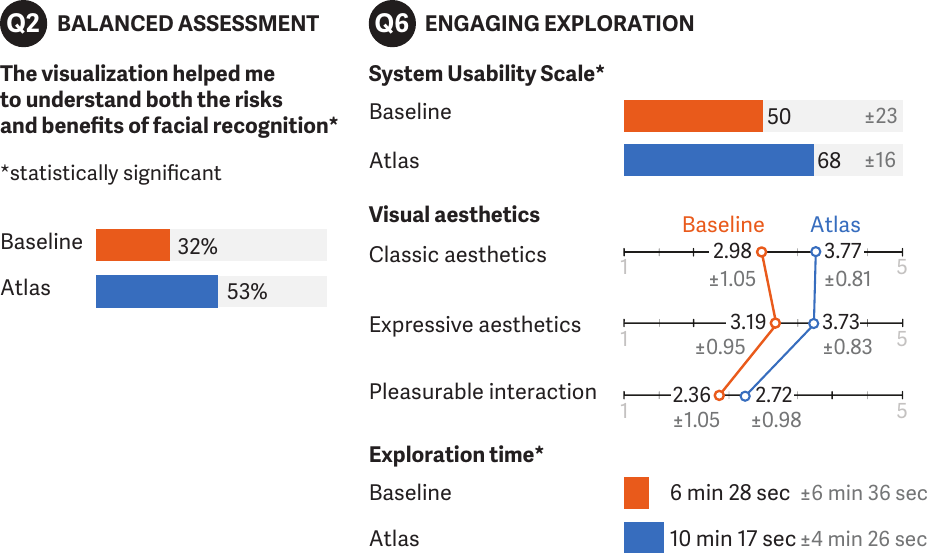}
  \caption{The Atlas outperformed baseline across all quantitative metrics. It offered a more balanced assessment of uses, scored higher in usability, visual aesthetics, and encouraged longer exploration time.}
  \label{fig:results}
\end{figure}

\subsection{Results}
We received a total of 140 responses, split equally into treatment and control (Supplementary Materials, Appendix E, Table \ref{tbl:participants_demographics}). Participants reflected the US population in terms of age, sex, and ethnicity.

\subsubsection{Quantitative results.}
\textbf{The Atlas provided a more balanced assessment of uses}, with 53\% of participants finding it offered comprehensive perspectives on risks, benefits, and mitigations, compared to 32\% of those who engaged with the baseline (Figure \ref{fig:results}Q2).

\begin{figure}[h!]
  \centering
  \includegraphics[width=0.55\textwidth]{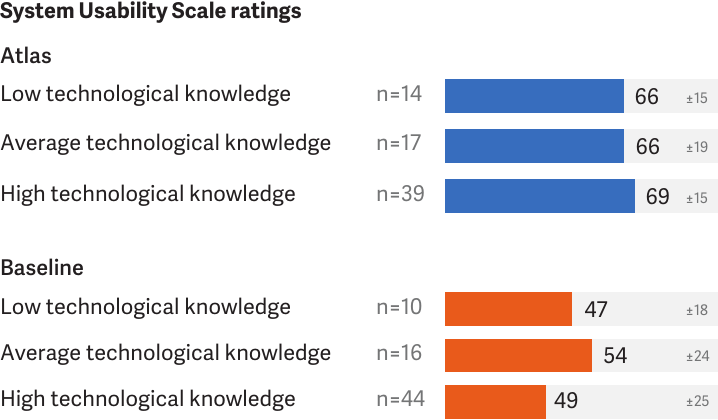}
  \caption{\textbf{System Usability Scale ratings by self-reported level of general technological knowledge.} The levels were assessed on a scale from 1 (definitely not above average) to 5 (definitely above average). Based on these assessments, users were categorized into three groups: those who identified as 1 or 2 were labeled as having \emph{low knowledge}, those who chose 3 were labeled as having \emph{average knowledge}, and those who chose 4 or 5 were labeled as having \emph{high knowledge} in technology.}
  \label{fig:sus}
\end{figure}

\vspace{2pt}
\noindent\textbf{Participants found the Atlas more usable than the baseline} with an average SUS score of 50, while the Atlas scored 68 (Figure \ref{fig:results}Q6). These higher scores were consistent across all levels of technological knowledge (Supplementary Materials, Appendix F, Figure \ref{fig:sus}). They also spent more time exploring the Atlas, indicating higher engagement.

\vspace{2pt}
\noindent\textbf{Participants found the Atlas more aesthetically pleasing than the baseline} and did so across all three dimensions of aesthetics (Figure \ref{fig:results}Q6). In \emph{classic aesthetics}, it provided a better-ordered structure. In \emph{expressive aesthetics}, it reflected more innovative way for presenting risks. In \emph{pleasurable interaction}, it was regarded as moderately successful, indicating room for improvement.

\subsubsection{Qualitative results.}
Participants (referred to as CP) found that the tool supported task completion by reducing the complexity of information about facial recognition ($n$=20), providing numerous examples of uses ($n$ = 18), listing their pros and cons ($n$ = 17), and effectively presenting them visually in the tool ($n$ = 17 mentions). As summarized by CP22, it \emph{``gave both quick tidbits, but also in depth explanations''}, while \emph{``making transitioning from one use to another feel smooth and understandable''} (CP26). The tool did not support participants who had strong pre-existing opinions about facial recognition ($n$ = 20) and needed more details to deliberate about each use ($n$ = 19). For example, CP9 felt they \emph{``didn't see much relating to environment''}, and CP6 wished for a feature where \emph{``AI can explain a picked dot like I'm 5''}.
\section{Demonstrating the Generalizability of the Tool}
To make sure the Atlas can handle any kind of AI, we use a straightforward five-part format. This format consists of:
\emph{purpose} (what the AI is meant to do), \emph{capability} (what the AI can actually do), \emph{AI user} (who uses the AI), \emph{AI subject} (on whom the AI works), and \emph{domain} (the area where the AI is used). This format is so flexible that it was first proposed to assess the risks of any AI system according to the EU AI Act \cite{Golpayegani2023Risk}.

\begin{figure}[h!]
  \centering
\includegraphics[width=\textwidth]{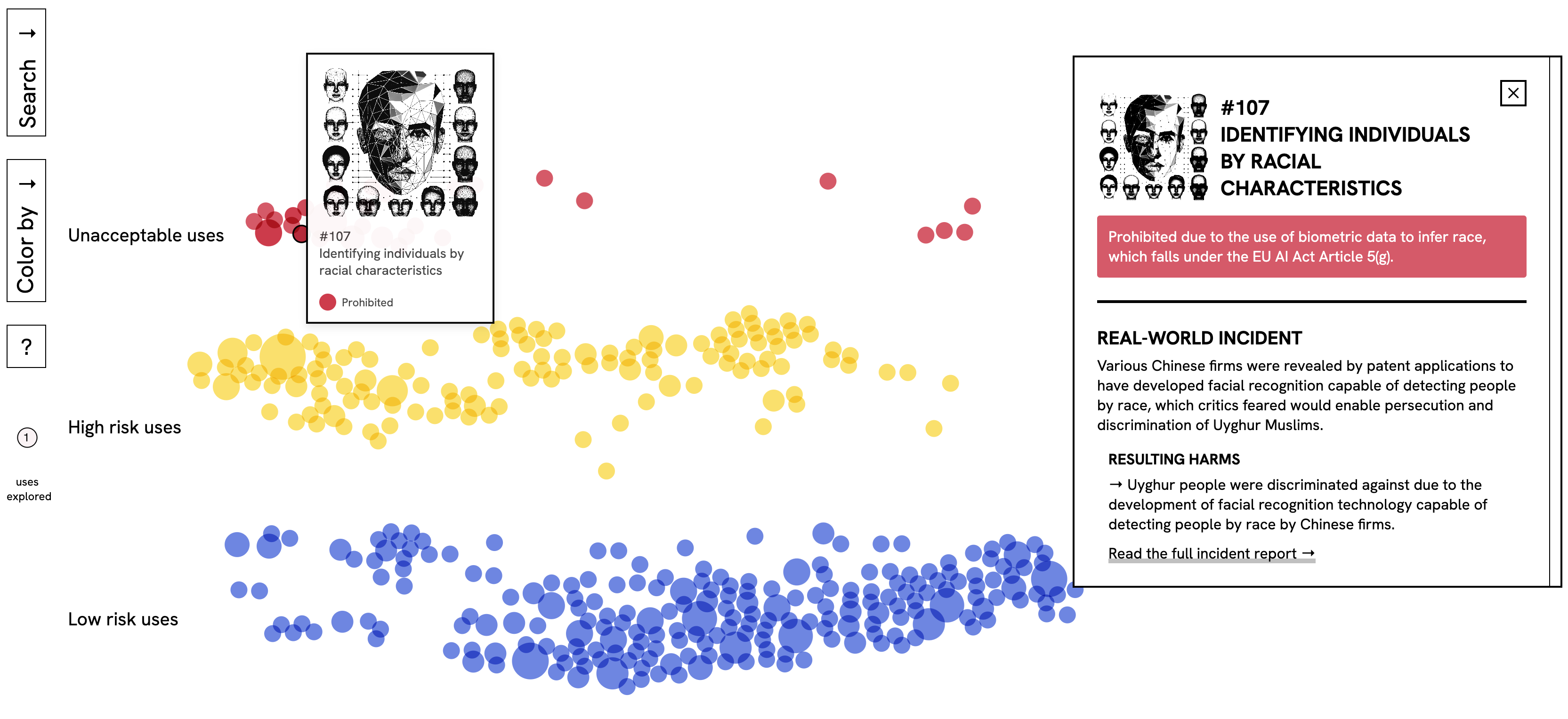}
  \caption{\textbf{Atlas of AI Risks for uses derived from the AI Incident Database.} We tested the generalizability of the Atlas by downloading 649 descriptions of AI-related incidents from the AI Incidents Database \cite{mcgregor2021preventing} and converting them into 379 specific AI uses. With multiple technologies, the spatial mapping adapted to present areas where their uses are similar.}
  \label{fig:incidents}
\end{figure}

\begin{figure}[h!]
  \centering
\includegraphics[width=\textwidth]{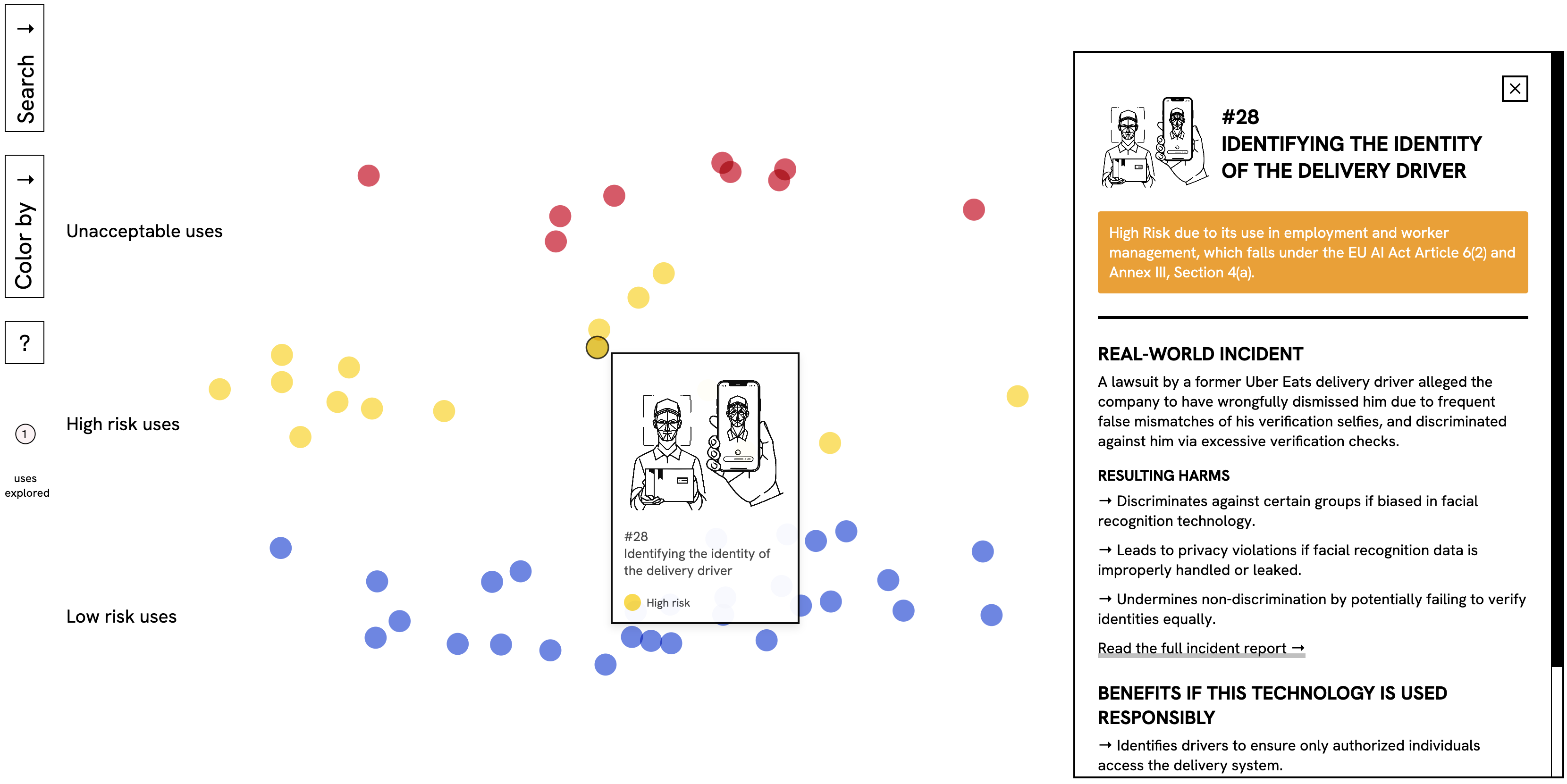}
  \caption{\textbf{Atlas of AI Risks for mobile computing uses derived from the AI Incident Database.} Drawing from the 379 specific AI uses identified in the database, we filtered 54 AI applications related to mobile computing by searching descriptions for mentions of mobile applications and technologies.}
  \label{fig:mobile}
\end{figure}

To then demonstrate the Atlas can handle any kind of AI  beyond facial recognition, we populated it with 379 real-world AI applications that resulted in news incident reports. We sourced the incident descriptions from the AI Incident Database (AIID), a standardized and verified collection of AI-related harmful events \cite{mcgregor2021preventing}. We populated the Atlas in five steps. First, we downloaded all 649 incident descriptions (e.g., ``YouTube's recommendation algorithms exposed children to disturbing videos''). Second, we used the \emph{ExploreGen} prompt to generate a possible AI use that could have caused the incident, breaking it down into the previously defined five components (e.g., ``Purpose'': ``Recommending suitable videos for children'', ``AI Capability'': ``Content filtering'', ``AI User'': ``YouTube'', ``AI Subject'': Children'', ``Domain'': ``Recommender Systems''). Third, we reviewed the formatted uses and merged the duplicate and most similar ones based on their semantic similarity and overlaps in our Atlas (e.g., 53 incident descriptions related to opearating autonomous taxis and delivery vehicles were merged into one use: ``Purpose'': ``Operating autonomus vehicles'', ``AI Capability'': ``Acting on sensor readings for navigation'', ``AI User'': Autonomous vehicle providers'', ``AI Subject'': ``Road users'', ``Domain'': ``Public and private transportation''). This review resulted in a final set of 379 uses. Fourth, we ran \emph{RiskGen}, \emph{BenefitGen}, \emph{MitigationGen}, and \emph{IllustrationGen} on each use to complete the content of the impact assessment card. Finally, we updated the Atlas's source code to include a dropdown menu, enabling users to explore all technologies from the incident database (Supplementary Materials, Appendix G, Figure \ref{fig:incidents}) or to focus on individual technologies like mobile computing (Supplementary Materials, Appendix G, Figure \ref{fig:mobile}). The extended Atlas is available online at~\url{https://social-dynamics.net/atlas}.

Based on the successful testing of its generalizability, we suggest three scenarios of how the Atlas could facilitate public debates and advocating for regulatory changes. First, by integrating the Atlas into AI city registers, where governments catalog urban technologies \cite{cityRegister}, citizens could better assess risks like privacy concerns before engaging in local discussions. Second, the Atlas could be integrated into consumer AI databases, such as the upcoming EU database for high-risk AI systems \cite{EUACT2024}, allowing individuals to evaluate AI products and make informed purchases. Finally, in education, the Atlas could serve students studying AI's ethical implications across different industries \cite{feffer2023ai}.

\section{Discussion and Conclusion}
\label{sec:discussion}

\noindent\textbf{Weighing Explanation and Exploration.} The goal of our tool is to effectively communicate AI uses and their risks to ordinary individuals interested in technology. This could also be achieved through other, more traditional methods, such as tutorials or training sessions. However, these approaches have two major limitations. First, they are often static and linear, limiting users' ability to explore content in a personalized way. Second, they demand significant time, which can overwhelm individuals with lower AI literacy. Instead, we developed a tool that aligns with the shift from static documentation, like impact assessment reports, to dynamic, glanceable formats like interactive model cards. By prioritizing visual narratives and aesthetic design, we encouraged users to spend more time exploring the tool and avoided overly technical designs that could alienate those who most need accessible learning tools.

\vspace{1pt}

\noindent\textbf{Future work.} We identify two areas for future work in data collection and visual presentation in similar tools. First is the exploration of complementary approaches that integrate collecting public perceptions and uses, expert assessments, and insights from LLMs. Second is the inclusion of visual features that help effectively build consensus around technology. These could include voting buttons where users can express their agreement or disagreement with certain assessments, and displaying aggregated community votes.

\vspace{1pt}

\noindent\textbf{Limitations.} This research comes with three limitations. First, LLM deployment faces issues like biases in training data that may overlook important risks and benefits \cite{luccioni2024stable}, while strict safety measures can omit risky uses already known to the public from past incidents \cite{AIRiskDatabase}. Addressing these issues requires fine-tuning models with specific technology datasets. For the Atlas, we addressed them by validating the content and manually removing any incoherent information.

Second, the Atlas usability score partly reflects the novelty and design challenge for similar tools and audiences. Our study's sample may not completely represent ordinary individuals interested in technology, due to limited controls over their location, age, ethnicity, gender, experience with AI and preferences in data visualization. Third, the study reflects US perspectives, which may not apply globally, as perceptions of technology risks and benefits vary by country. For example, facial recognition is most accepted in China, less so in Germany, with the US and UK in between \cite{frtPerceptions2021}. While most Chinese citizens emphasize benefits like convenience. Americans, Germans, and Brits are more concerned about risks such as surveillance \cite{frtCrossCountryPerceptions_2023}. Therefore, our findings should be interpreted with these limitations in mind.

Our work initially focused on designing and evaluating a tool for ordinary individuals interested in technology. However, we found that even non-tech savvy users could effectively engage with it when prompted, as the tool’s usability remained consistent across users with different levels of technological knowledge. This demonstrates the Atlas's ability to make complex AI technology risks understandable, filling the gap in current risk communication methods.

\bibliography{aaai24}
\clearpage
\onecolumn
\section{Supplementary Materials}

\subsection{Appendix A \\Summary of All Technology Uses, Risks, Mitigations and Benefits Generated During the Formative Study}

\begin{figure*}[h!]
  \centering
  \includegraphics[width=0.95\textwidth]{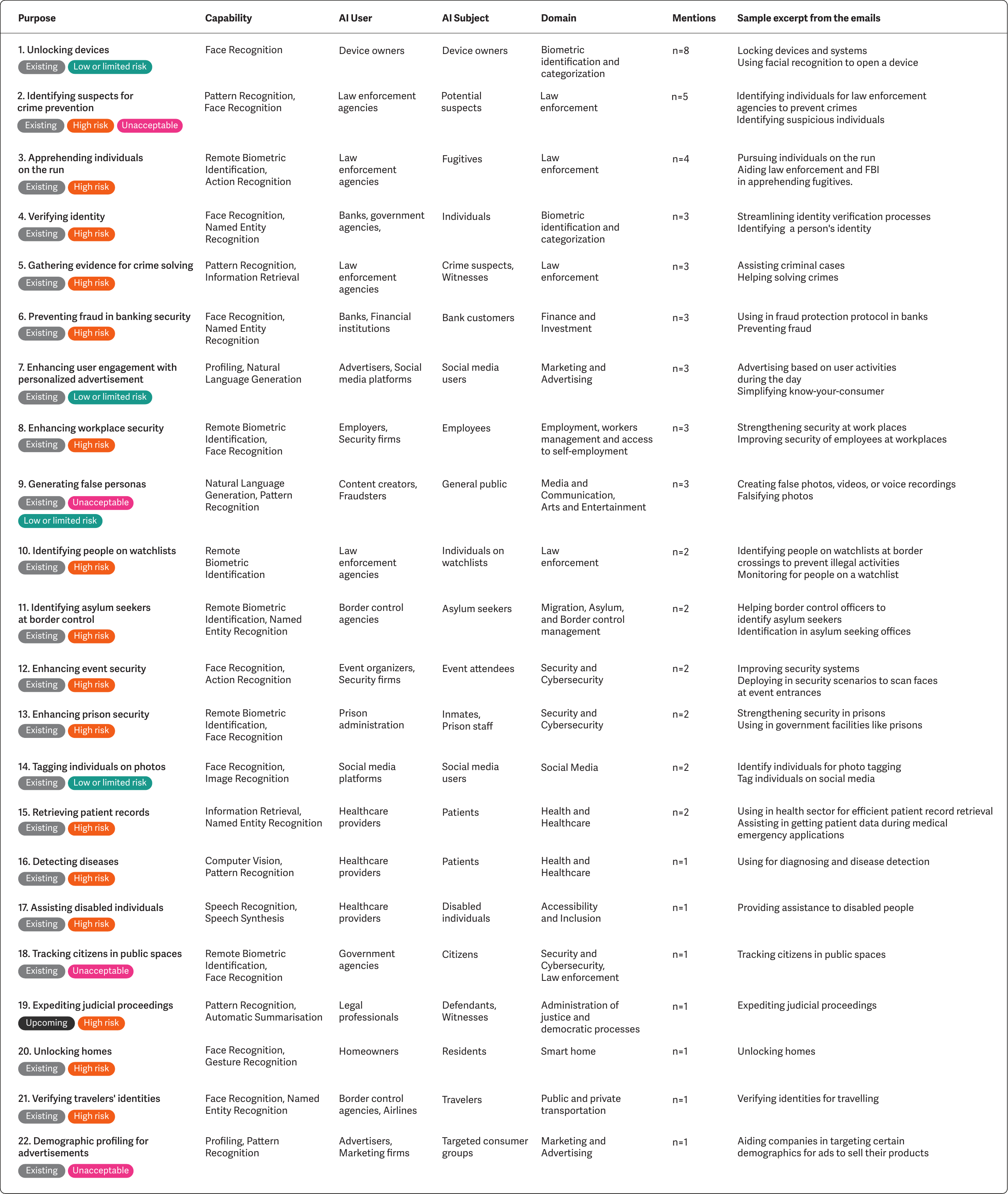}
  \caption{Uses generated by the participants of the formative study through writing emails to regulators.}
  \label{fig:uses}
\end{figure*}

\begin{figure*}[h!]
  \centering
  \includegraphics[width=0.95\textwidth]{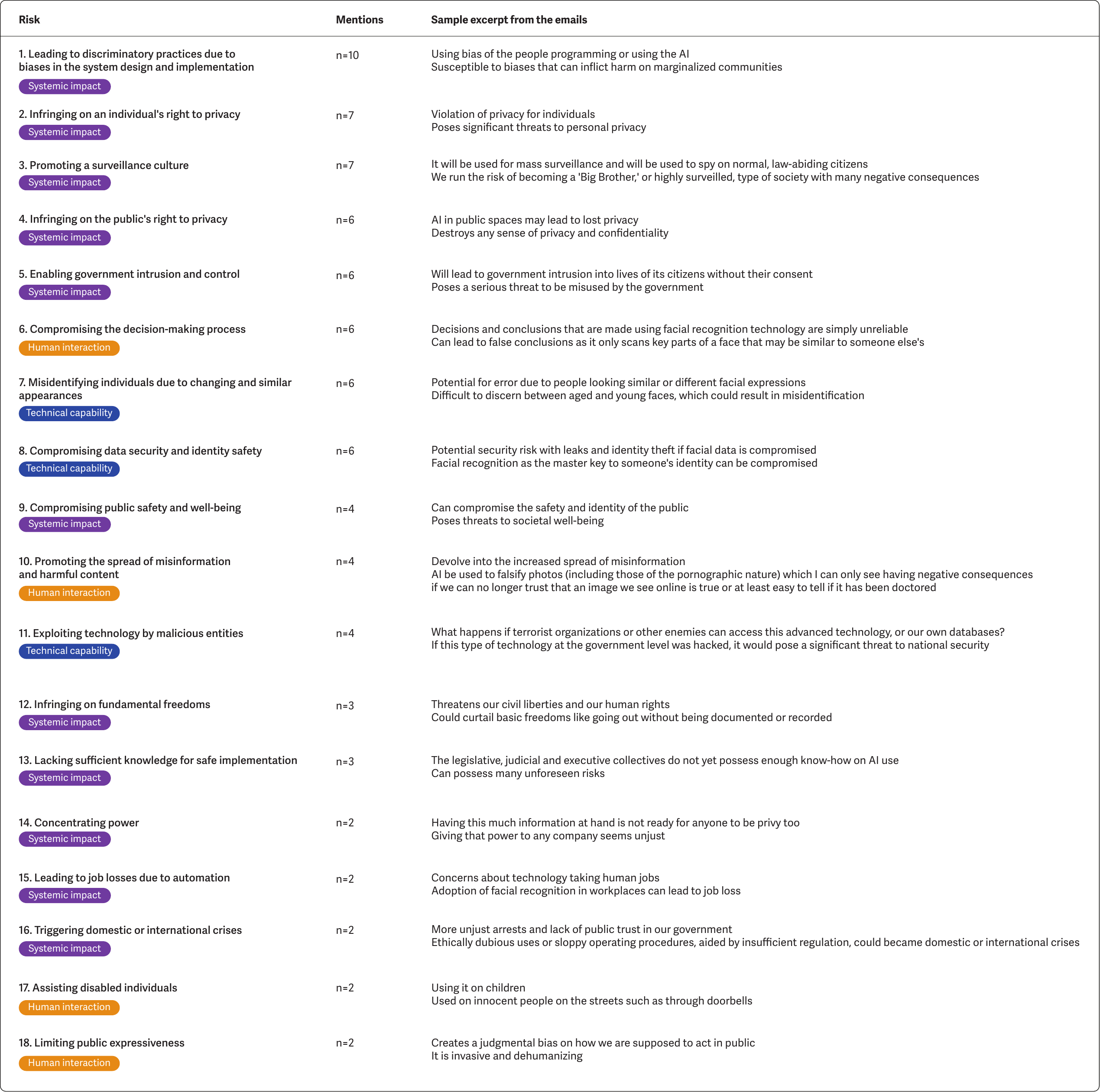}
  \caption{Risks generated by the participants of the formative study through writing emails to regulators.}
  \label{fig:risks}
\end{figure*}

\begin{figure*}[h!]
  \centering
  \includegraphics[width=0.9\textwidth]{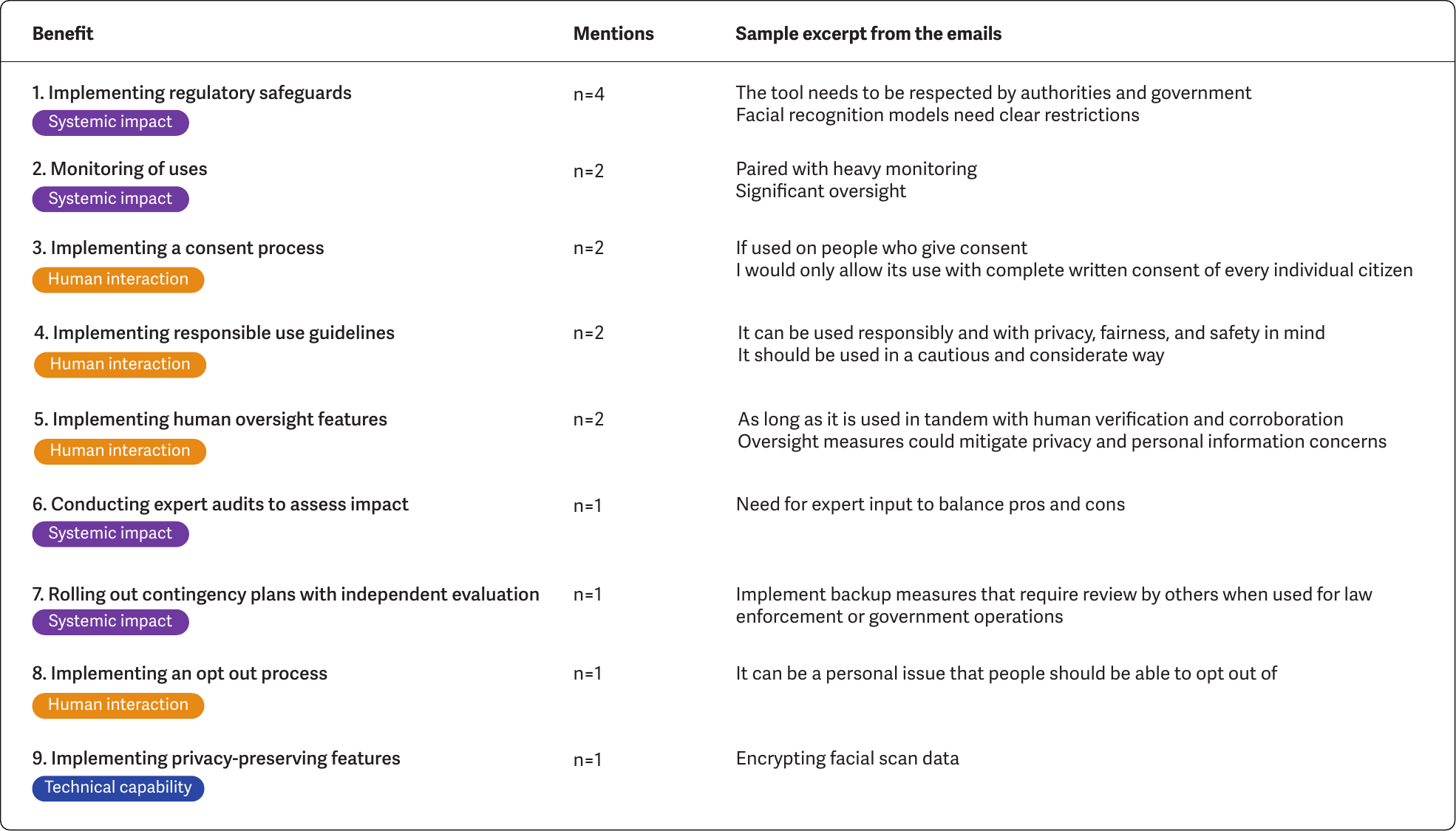}
  \caption{Mitigations generated by the participants of the formative study through writing emails to regulators.}
  \label{fig:mitigations}
\end{figure*}

\begin{figure*}[h!]
  \centering
  \includegraphics[width=0.9\textwidth]{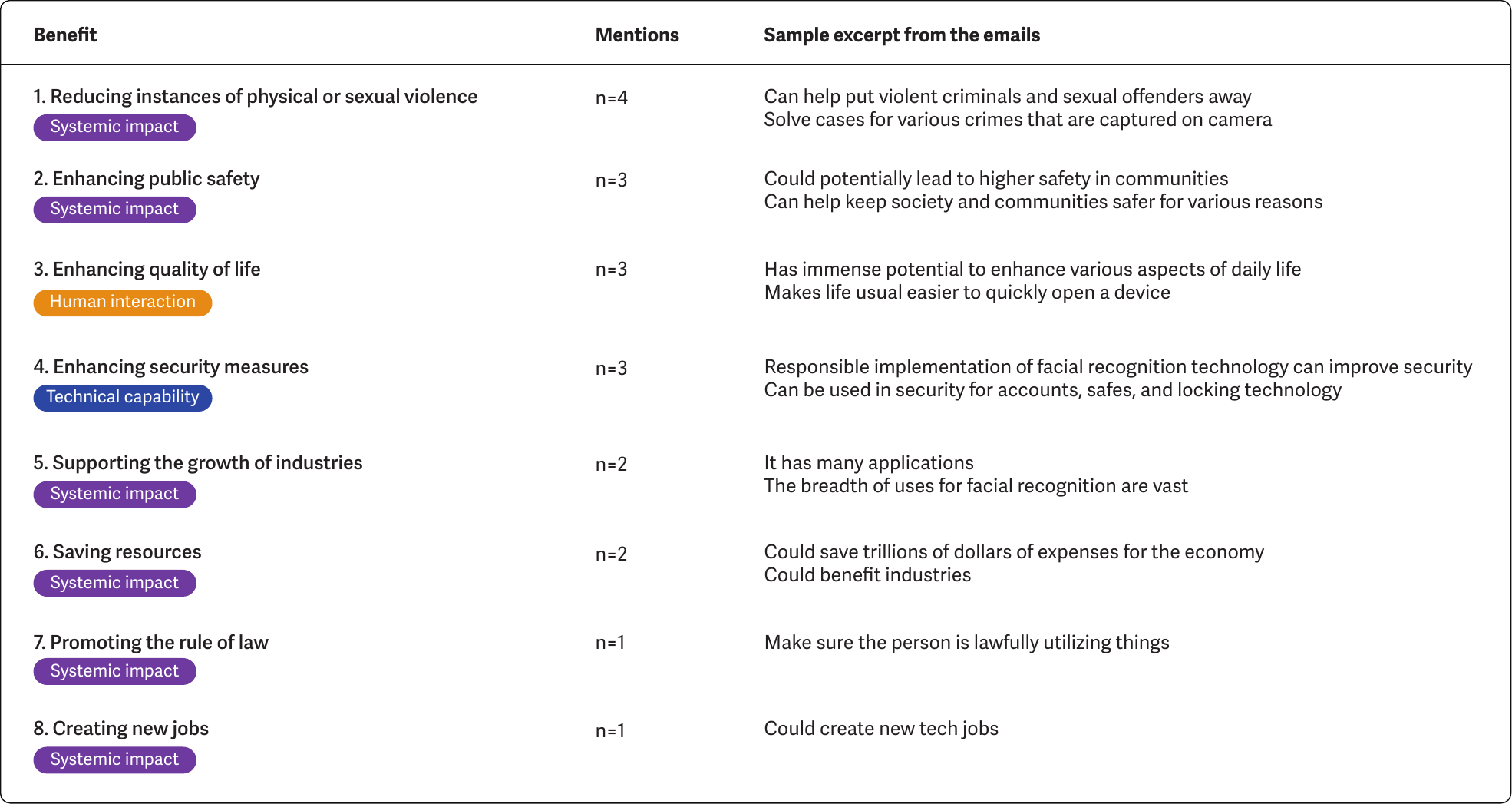}
  \caption{Benefits generated by the participants of the formative study through writing emails to regulators.}
  \label{fig:benefits}
\end{figure*}

\clearpage
\subsection{Appendix B \\Prompt Engineering for Generating Facial Recognition Technology Uses, Along With Their Risks, Mitigations, Benefits, and Illustrations}
The data for our Atlas was generated using five adaptable prompts: \emph{ExploreGen} \cite{herdel2024exploregen}, \emph{RiskGen}, \emph{BenefitGen}, \emph{MitigationGen} \cite{constantinides2024_risks_benefits, AIDesign2024} and \emph{IllustrationGen}. To execute them, we chose OpenAI's GPT-4 model for its leading performance as of June 2024 \cite{leaderboard}, and the DALL-E 3 model for its good integration with GPT-4 and good image quality. 

\smallskip
\textbf{(1) ExploreGen} prompt generates a list of existing and upcoming uses of a given technology (i.e., facial recognition) across various application domains \cite{herdel2024exploregen}. The prompt includes four key elements to generate diverse and realistic uses: \emph{role}, \emph{domains}, \emph{instructions}, and \emph{output}. 

First, we asked LLM to take the \emph{role} of a ``Senior AI Technology Expert responsible for identifying and cataloging various AI applications and use cases''. This role draws on the LLM's prior knowledge and ensures generation of realistic AI uses. 

Second, we directed the LLM to generate three distinct uses for each of the \emph{46 domains} that cover key economic sectors and various aspects of everyday life, ensuring diverse AI uses.

Third, we formulated \emph{instructions} to break down the uses into five components suitable for risk assessment, aligning with the requirements of the EU AI Act \cite{Golpayegani2023Risk}. These components are: \textit{domain} (the specific industry or sector, e.g., \emph{finance}), \textit{purpose} (the end goal, e.g., \emph{fraud prevention}), \textit{capability} (the technological feature, e.g., \emph{facial recognition}), \textit{AI user} (the entity managing and overseeing the user, e.g., \emph{banks}), and \textit{AI subject} (individuals or groups impacted by the use, e.g., \emph{customers}).

Finally, the \emph{output} of \emph{ExploreGen} is a list of facial recognition uses. Each use is described in a five-component format (e.g., [finance, fraud prevention, facial recognition, banks, customers]]) and is summarized in concise one-line description (e.g., facial recognition for financial fraud detection).

\smallskip
\noindent\textbf{(2) RiskGen} prompt generates and categorizes potential risks for each use, focusing on issues such as infringing on Human Rights (HRs) \cite{rights1961universal}, hindering Sustainable Development Goals (SDGs) \cite{sdgs}, and failing to adhere to regulatory guidelines from the EU AI Act \cite{EUACT2024}. Together with \emph{BenefitGen} that will be discussed below, \emph{RiskGen} helps to categorize uses and provide a balanced assessment of their risks and benefits. \emph{RiskGen} consists of four elements: \emph{role}, \emph{input}, \emph{instructions}, and \emph{output}. 

First, the \emph{role} is a ``Senior AI Technology Expert, specializing in compliance with the EU AI Act, SDGs, and HRs''.

Second, the \emph{input} includes excerpts from the EU AI Act which describe high-risk AI systems (e.g., Annex III), the 17 SDGs definitions from the Sustainable Development Agenda \cite{sdgs}, and the 30 HRs articles from the Universal Declaration of Human Rights (UDHR) \cite{rights1961universal}.

Third, the \emph{instructions} use a Chain-of-Thought approach, i.e., dividing the task into a series of smaller, intermediate reasoning steps that lead to the final output~\cite{NEURIPS2022_9d560961}. These steps are: (1) producing a risk classification as per the EU AI Act (whether the use is unacceptable, high-risk, or none of these two (minimal risk)); (2) identifying any additional risks from the ways the use undermines SDGs or HRs; and (3) grouping all risks based on their relevance to the capability, human interaction, and systemic impact layers \cite{weidinger2023sociotechnical}. 

Fourth, the \emph{output} of the prompt includes the risk classification, completed with LLM's reasoning for this classification and relevant excerpts from the EU AI Act. This structure is similarly repeated for HRs and SDG risks, which include excerpts from relevant UDHR Articles or SDGs. The output of \emph{RiskGen} is further passed to the \emph{MitigationGen}.

\smallskip
\noindent\textbf{(3) BenefitGen} prompt generates and categorizes potential benefits for each use. It consists of four elements: \emph{role}, \emph{input}, \emph{instructions}, and \emph{output}. Similarly to \emph{RiskGen}, the \emph{role} is a ``Senior AI Technology Expert, specializing in SDGs and HRs'', and the \emph{input} includes the 17 SDGs definitions the 30 HRs articles. The \emph{instructions} and \emph{output} also mimic those of \emph{RiskGen}. 
Two primary differences are that we ask LLM to generate \emph{benefits} instead of \emph{risks} for HRs and SDGs, and we remove any references to the EU AI Act, as it is a risk-based regulation that may not be applicable when generating benefits.

\smallskip
\noindent\textbf{(4) MitigationGen} prompt proposes mitigation strategies for each of the risks previously identified by \emph{RiskGen}. It consists of three elements: \emph{role}, \emph{instructions}, and \emph{output}. 

Second, the \emph{instructions} include four steps: (1) proposing mitigation strategies; (2) grouping these strategies based on their relevance to the capability, human interaction, and systemic impact layers \cite{weidinger2023sociotechnical}, (3) generating a description of a new mitigated version of the use; and (4) evaluating the compliance of the mitigated version with the EU AI Act (assessing whether that version is still unacceptable or high-risk).

Third, the \emph{output} contains mitigation strategies, their groups, description of the new mitigated system, and its EU AI Act classification. 

\smallskip
\textbf{(5) IllustrationGen} illustrates the uses to fulfil design requirement \emph{(R5)}. This is achieved with a prompt asking to \emph{``Generate an image for the [description of the use] with the content that is safe and appropriate. Use line art style, low polygons, and black lines on the white background''}. We sent a prompt to OpenAI's DALL-E 3 model, which generated images with a resolution of 1024x1024 pixels each.

\clearpage
\subsection{Appendix C \\The Setup of the User Study Evaluating the Atlas}

\begin{figure}[ht]
  \centering
\includegraphics[width=\textwidth]{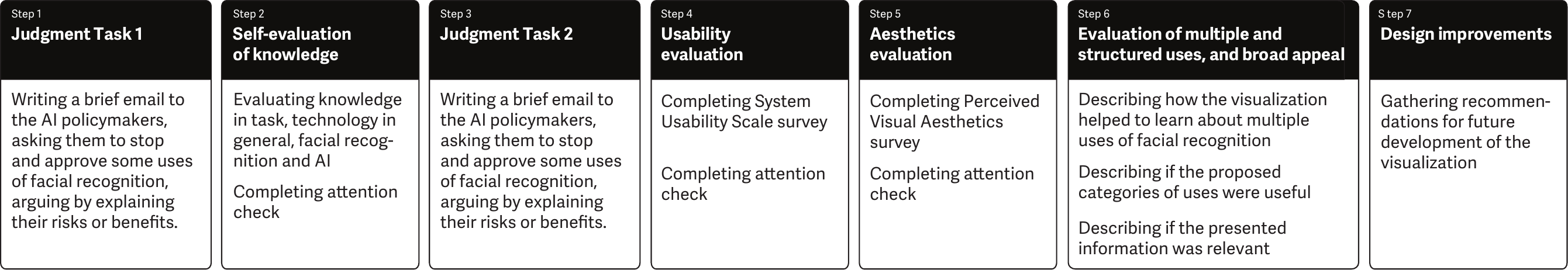}
  \caption{\textbf{The evaluation study involved seven steps.} As a first step, participants completed the first task without tools (Step 1). Next, they answered control questions and an attention-check (Step 2). Subsequently, they used our tool (treatment) or a baseline (control) for analysis and a second judgment task (Step 3), followed by usability evaluation and another attention-check (Step 4). Next, they assesses the visual aesthetics of the tool and completed a third attention-check (Step 5). Then they described how the visualization helped them lo learn about multiple uses, if their categorization was useful and if the overall presented information was relevant (Step 6). The final step gathered recommendations for future tool development (Step 7). }
  \label{fig:survey_setup}
\end{figure}

\subsection{Appendix D \\Baseline Visualization of AI Use Risks}

\clearpage
\subsection{Appendix E \\Demographic Details of Participants of the User Study Evaluating the Atlas}

\begin{table}[h!]
\centering
\renewcommand{\arraystretch}{1.1} 
\setlength{\tabcolsep}{3.25pt}
\caption{Self-reported knowledge and demographics of ordinary individuals interacting with either the baseline or the Atlas.}
\label{tbl:participants_demographics}
\begin{tabular}{|p{4.8cm}|p{6cm}|p{1.6cm}|p{1.6cm}|p{1.8cm}|}
\hline
\textbf{Control} & \textbf{Characteristic} & \begin{tabular}[c]{@{}l@{}} \textbf{Baseline}\\ (n=70)\end{tabular} & \begin{tabular}[c]{@{}l@{}} \textbf{Atlas}\\ (n=70) \end{tabular} & \begin{tabular}[c]{@{}l@{}} \textbf{US Census} \\ (\citeyear{census_ethnicity_2020, census_age_gender_2022}) \end{tabular} \\ \hline

\multirow{4}{*}{\begin{tabular}[c]{@{}l@{}} \textbf{Knowledge} \\ (1: definitely not above average, \\ 5: definitely above average) \end{tabular}}       
                            & Task (writing an email to policymakers)  & 3.09 & 2.96 & - \\ \cline{2-5}
                            & Technology in general & 3.53 & 3.46 & - \\ \cline{2-5}
                            & Facial recognition & 2.87 & 2.73 & - \\ \cline{2-5}
                            & Artificial Intelligence & 3.16 & 3.00 & - \\ \hline

\multirow{5}{*}{\begin{tabular}[c]{@{}l@{}} \textbf{Age} \\ (\% of population above 18 years) \end{tabular}}    
                            & 18-29 & 21\% & 20\% & 20\% \\ \cline{2-5}
                            & 30-39 & 19\% & 20\% & 18\% \\ \cline{2-5}
                            & 40-49 & 17\% & 17\% & 16\% \\ \cline{2-5}
                            & 50-59 & 17\% & 16\% & 16\% \\ \cline{2-5}
                            & 60 and above & 26\% & 27\% & 30\% \\ \hline

\multirow{2}{*}{\begin{tabular}[c]{@{}l@{}} \textbf{Sex} \\ (\% of population) \end{tabular}}         
                            & Female & 51\% & 49\% & 50\% \\ \cline{2-5}
                            & Male & 49\% & 51\% & 50\% \\ \hline

\multirow{6}{*}{\begin{tabular}[c]{@{}l@{}} \textbf{Ethnicity} \\ (\% of population) \end{tabular}}      
                            & White & 60\% & 61\% & 62\% \\ \cline{2-5}
                            & Black & 14\% & 11\% & 12\% \\ \cline{2-5}
                            & Asian & 7\% & 7\% & 6\% \\ \cline{2-5}
                            & Mixed & 10\% & 10\% & 10\% \\ \cline{2-5}
                            & Native American or Alaskan Native & 1\% & 1\% & 1\% \\ \cline{2-5}
                            & Other & 7\% & 9\% & 9\% \\ \hline

\end{tabular}
\end{table}

\subsection{Appendix F \\System Usability Scale Ratings by Self-Reported Level of General Technological Knowledge}

\clearpage
\subsection{Appendix G \\Atlas Views Generalized for the Uses From the AI Incident Database}

\end{document}